\documentclass[12pt]{iopart} 
\usepackage{graphicx,iopams}
\usepackage{subfigure}
\usepackage{dsfont}

\newcommand{\be}{\begin{equation}}
\newcommand{\ee}{\end{equation}}
\newcommand{\bea}{\begin{eqnarray}}
\newcommand{\eea}{\end{eqnarray}}
\newcommand{\beaa}{\begin{eqnarray*}}
\newcommand{\eeaa}{\end{eqnarray*}}

\newcommand{\nn}{\nonumber}
\def\rf#1{(\ref{#1})}

\begin{document} 
\title{An analysis of the phase space of Ho\v{r}ava-Lifshitz cosmologies} 
\author{S Carloni\dag, E Elizalde\ddag, P J Silva\P}

\address{\dag Institut d'Estudis Espacials de Catalunya
(IEEC) \\
Campus UAB, Facultat de Ci\`encies, Torre C5-Par-2a pl \\ E-08193 Bellaterra
(Barcelona) Spain} 
\address{\ddag Consejo Superior de Investigaciones Cient\'{\i}ficas (ICE/CSIC) \, and Institut d'Estudis
Espacials de Catalunya (IEEC), Campus UAB, Facultat de Ci\`encies, Torre C5-Par-2a pl, E-08193 Bellaterra
(Barcelona) Spain} 
\address{\P Institut d'Estudis Espacials de Catalunya
(IEEC) and Institut de F\'{\i}sica d'Altes Energies,
Campus UAB, Facultat de Ci\`encies, Torre C5-Par-2a pl, E-08193 Bellaterra
(Barcelona) Spain} 
\ead{carloni@ieec.uab.es,elizalde@ieec.fcr.es,psilva@ifae.es} 
\begin{abstract} 
Using  the dynamical system approach, properties of cosmological models based on the
Ho\v{r}ava-Lifshitz  gravity are systematically studied. In particular, the cosmological
phase space of the Ho\v{r}ava-Lifshitz model is characterized. The analysis allows to compare some key
physical consequences of the imposition (or not) of detailed balance. A result of the
investigation is that in the detailed balance case one of the attractors in the theory corresponds
to an oscillatory behavior. Such oscillations can be associated to a bouncing universe,
as previously described by Brandenberger, and will prevent
a possible evolution towards a de Sitter universe. Other results obtained show that the cosmological models generated by Ho\v{r}ava-Lifshitz gravity without the detailed balance assumption have indeed the potential to describe the transition between the Friedmann and the dark energy eras. The whole analysis leads to the plausible conclusion that a cosmology compatible with the present observations of the universe can be achieved only if the detailed balance condition is broken.
\end{abstract} 


\maketitle 
\section{Introduction}

Recently, Ho\v{r}ava made a proposal for an ultraviolet completion of general relativity (GR) normally referred to as the Ho\v{r}ava-Lifshitz (HL) theory \cite{h}. (The name is due to Ho\v{r}ava's initial inspiration on the Lifshitz theory in solid state physics.) The salient characteristic of the HL proposal is that it seems to be renormalizable, at least at the level of power counting. This ultraviolet behavior is obtained by introducing irrelevant operators that explicitly break Lorentz invariance but ameliorate the ultraviolet divergences. On the other hand,  Lorentz invariance is expected to be recovered at low energies, as an accidental symmetry of the theory.

Originally, the HL proposal came with the possibility of imposing or not the so-called {\it projectability condition} and the {\it detailed balance condition}. The first condition is related to the space-time dependence of the lapse function, $N$, which characterizes a canonical $3+1$ decomposition of the metric field $g$, while the second is a restriction on the form of the potential terms which may appear in the Lagrangian that leads to simplifications since it reduces the final number of couplings. Notice therefore that we have, in principle, four different incarnations of this proposal.

Since its publication, the HL theory has been the object of an exhaustive research regarding its different properties and implications to space-time physics. In particular, a lot of attention has been paid to its internal consistency, to how to define the infrared limit, its compatibility with GR, and the potential application of the results obtained to cosmology.

Presently---as this article is being written---the consistency status of the theory is not completely clear, nor its low energy limit and, hence, how GR is recovered at the different regimes. In fact, for the non-projectable version, although the low energy limit has been found \cite{afsh}, there seem to remain important problems, regarding strongly coupled features that may preclude any type of perturbation approach \cite{strongcoupling,Blas:2009yd}, while the corresponding symplectic structure and systems of constraints do not seem to be well defined \cite{Kobakhidze:2009zr,Kocharyan:2009te,Mukohyama:2009mz}. Also, imposing detailed balance leads to a cosmological constant with the {\it wrong sign}, from what we should expect to be at odds with cosmological observations \cite{h,Kiritsis:2009sh}. In comparison, the projectable version seems to be less problematic, since the above listed problems can in principle be evaded by the non-local form of the Hamiltonian constraint \cite{Kocharyan:2009te}.
Also, if detail balance is not imposed a richer phenomenology seems to appear, where cosmological applications may lead to new results in inflation, bouncing cosmology, dark matter, and dark energy (see, for example, \cite{Kiritsis:2009sh,Sotiriou:2009bx,Sotiriou:2009gy,Nastase:2009nk,Brandenberger:2009yt,Wang:2009rw,Calcagni:2009ar,Piao:2009ax,Park:2009zr-zra,Appignani:2009dy,Cai:2009in}).

At this point it is important to investigate the key aspects of the theory, which may help in clarifying the status of the different HL proposals as plausible candidates of a quantum theory of gravity. By now, it is clear that the study of the consistency of the theory is a complicated problem that will still take a lot of effort and time to ascertain and that will involve detailed research of the renormalization properties of the theory and, most probably, of its non-perturbative nature (see e.g. \cite{Orlando:2009az,Visser:2009fg}). On the other hand, another important requirement the theory should fulfill is its implication with phenomenology and, in particular, with cosmology. In this last regard there are already papers on some of the consequences of these new scenarios\footnote{See \cite{Wang:2009rw,Leon:2009rc} for some general results of this type based on different assumptions and technics.}; however, no systematic study of the possible form of the cosmological relevant solutions of the theory, either for the detailed balance case or for a general potential, has appeared yet.

In the present work, motivated by the above facts, we investigate the space of all possible classical solutions of the HL theory that are relevant for cosmology, i.e. we study the {\it cosmological phase space} of the HL model. To address this involved issue, we will borrow techniques from the field of dynamical systems that are frequently used on more canonical studies applied to diverse types of cosmologies. As a matter of fact, dynamical system techniques in cosmology have been used for many years. In our work, we will consider the dynamical system approach to cosmology as formulated by Collins and applied by Wainwright and Ellis \cite{ellisbook}. Such approach has the advantage of relying on dynamical system variables which are directly related to cosmological observables---like the matter density parameter---and of being relatively easy to apply to very complicated cosmological models. During the past years this method has been able to unfold some very interesting properties of Bianchi universes \cite{ellisbook}, as well as of scalar-tensor and higher-order gravity cosmological models (for some detailed examples, see \cite{Carloni:2007eu,Carloni:2004kp,Carloni:2007br,shosho}).

As we will see, when applied to HL cosmologies these techniques will be able to reveal some interesting results. In particular, when detailed balance is imposed, although one is able to prove that the corresponding cosmology could---given suitable initial conditions---behave like a Friedmann one, it turns out that it will eventually evolve towards an oscillatory behavior or either recollapse, leaving no space for a dark-energy era. When detailed balance is broken, however, the situation changes: given the proper initial conditions and values of the parameters the cosmology undergoes an almost Friedmann era(s) to evolve towards a de Sitter era which can be associated to dark energy. In this sense the plausible consequence is that a cosmology compatible with the present observations of the universe can be achieved {\it only} if the detailed balance condition is broken. In addition to that, we will show that the information on the cosmological relevance of the values and signs of the parameters in the theory helps to uncover properties of the general theory. For example, from the relative sign of the Hubble and matter terms in the cosmological equations, we will be able to deduce that the spin zero modes of the theory can be excluded.

The paper is organized as follows. In Sect.~\ref{1} we present the Lagrangian of the HL theory for the two cases of either imposing or not the detailed balance constraint, where the minimal potential defined in \cite{Sotiriou:2009bx} for the non-detailed-balance case will be used. The projectability condition does not affect our analysis, since for the cosmological {\it ansatz} the shift $N$ is only time dependent. In Sects.~\ref{2} and \ref{3} we present the specific analysis for the theory with detailed balance and for the one without detailed balance, respectively. We will be able to characterize the classical phase space, discussing its structure in depth, in particular all the fixed points and their nature as repellers or attractors, in each of the cases. Finally, in Sect.~\ref{4} we summarize the results obtained, giving some perspectives for further work.

\section{ Ho\v{r}ava-Lifshitz gravity}
\label{1}

The following is a short overview on the minimal set of definitions needed to understand the dynamical system analysis of the forthcoming sections. In the HL theory, the dynamical variables are defined to be the laps $N$, the shift $N_i$ and the space metric $g_{\,ij}$, Latin indices running from 0 to 3. The space-time metric is defined using the ADM construction, as
\bea
ds^2=-N^2dt^2+g_{ij}(dx^i+N^idt)(dx^j+N^jdt)\;,
\eea
where $N^i=g^{ij}N_j$ as usual.
The action $S$ is written in terms of geometric objects, characteristics of the ADM slicing of space-time, like the 3d-covariant derivative $\nabla_i$, the spatial curvature tensor $R_{ijkl}$, and the extrinsic curvature $K_{ij}$. They are defined as follows,
\begin{equation}
R^{i}{}_{jkl}=W^i{}_{jl,k}-W^i{}_{jk,l}+ W^m{}_{jl}W^i{}_{km}-
W^n{}_{jk}W^i{}_{lm}\;,
\end{equation}
where $W^i{}_{jl}$ are the Christoffel symbols (symmetric in the lower indices), given by
\begin{equation}
W^i_{jl}=\frac{1}{2}g^{im}
\left(g_{jm,l}+g_{ml,j}-g_{jl,m}\right)\,.
\end{equation}
The Ricci tensor is obtained by contracting the {\em first} and the
{\em third} indices
\begin{equation}\label{Ricci}
R_{ij}=g^{kl}R_{ikjl}\quad\hbox{and}\quad R=g^{ij}R_{ij}\,,
\end{equation}
while the extrinsic curvature is defined as
\be
K_{ij}={1\over2N}(-\dot{g}_{ij}+\nabla_iN_j+\nabla_jN_i)\,,
\ee
where the dot stands for time derivative.

In terms of the above tensor fields, the HL action can be written as
\bea
S=\int dt\, dx^3\,N\sqrt{g}\left({\cal L}_{kinetic}-{\cal L}_{potential}+{\cal L}_{matter}\right)\,,
\eea
being the kinetic term universally given by
\be
\label{k}
{\cal L}_{kinetic}=\alpha(K_{ij}K^{ij}-\lambda K^2)\,,
\ee
with $\alpha$ and $\lambda$ playing the role of coupling constants. The potential term is, in principle, a generic function of $R_{ijkl}$ and $\nabla_i$. Here we will work with two different types of potentials: first, the detailed balance potential and, second, the potential defined in \cite{Sotiriou:2009bx} (the SVW case), that corresponds to the more general potential with dimensionless kinetic couplings (see the original article for clarifications.). Finally, the matter term corresponds to the coupling of the matter fields to gravity.

The potential for the detailed balance case is
\be
{\cal L}_{potential-detail}=\beta C_{ij}C^{ij}+\gamma \epsilon^{ijk}R_{il}\nabla_j R^l_k + \zeta R_{ij}R^{ij}+\eta R^2+\delta R+ \sigma\,,
\ee
where $\sqrt{g}C^{ij}=\epsilon^{ikl}\nabla_k(R^j_l-1/4Rg^j_l)$ and the Greek letters define the different coupling constants. We have closely followed the notation of \cite{Kiritsis:2009sh} for simplicity. We stress that only three out of the whole set of coupling constants above are independent, due to the detailed balance constraint.  Then, it is not difficult to see that, if we want the theory to make contact with GR in the infrared regime, we have to do the following identifications:
\be
\label{ir}
\alpha={c\over 16\pi G_N}\,,\quad \lambda= 1\,,\quad  \delta={1\over 16\pi c G_N}\,,\quad \sigma={-2\Lambda\over 16\pi c G_N}\,,
\ee
with all other couplings being equal to zero. To get the above results, we also have defined a new time coordinate $x^0=ct$, where $(c,G_N,\Lambda)$ are the velocity of light, Newton's constant, and the cosmological constant, respectively. From this point onwards, we will keep $c=1$, so that our time coordinate is actually $t$.

The potential of the SVW case is
\bea
\fl{\cal L}_{potential-SVW}&=&g_8\nabla_iR_{jk}\nabla^iR^{jk}+g_7R\nabla^2R +g_6R^i_jR^j_kR^k_i+g_5R(R_{jk}R^{jk}) +g_4R^3\\ \nonumber
&&+g_3R_{jk}R^{jk}+g_2R^2+g_1R+g_0\,,
\eea
where we have modified the original notation to better accommodate our formulation of the universal kinetic term \rf{k}, therefore our coupling constants are not all dimensionless. As in the detailed balance case, in order to recover GR at low energies we get similar equations to those in \rf{ir}. Just notice we need to identify $\delta\Leftrightarrow g_1$ and $\sigma\Leftrightarrow g_0$, while all the other couplings in the potential should go to zero.

At this point, other phenomenological and theoretical considerations may help to constraint the range of values that the different couplings should take. For example, in \cite{Bogdanos:2009uj} it was found that ghost instabilities are present if $\lambda\in (1/3,1)$, the cosmological constant is negative for the detailed balance potential, $\alpha > 0$, etc. Here, we will constraint as little as possible the different ranges of values on each coupling constant to see how much information comes out of the dynamical system approach. Then, we will add this information to the constraints arising from other considerations, to finally obtain the most promising form of the potential. In particular, we will take $\lambda$ different from 1/3 (corresponding to the scale invariant case) as the only limitation on its range.

To start our studies on cosmological applications, we have to consider a FLRW {\it ansatz} on the 4D metric. Due to the homogeneity and simple time dependence of this type of {\it ansatz}, only a subset of the coupling constants plays a role in the dynamics, in any case. Also, as we already said, $N$ is a function of time only and, therefore, all issues related to the projectability condition turn out to be irrelevant\footnote{Recall that we are interested in the classical phase space and its structure. On the other hand, the dynamics of fields propagating in the above solutions are determined by the specific incarnation of the HL theory chosen, where projectability is important.}. Then, we set,
\bea
N\longrightarrow N(t)\,,\quad N_i\longrightarrow 0\,,\quad g_{ij}\longrightarrow a(t)\gamma_{ij}\,,
\eea
where $\gamma_{ij}$ is a maximally symmetric 3D metric, of constant curvature $R=6k$, with $k=(-1,0,1)$.

The inclusion of matter content in the theory in its general form has not been worked out yet. There are various discussions and suggestions (see for example \cite{Kiritsis:2009sh}), but this is still an open subject. An interesting approach is based on adding some new degrees of freedom to the framework, so that the theory gains in diffeomorphism invariance. Then, minimal coupling to matter is invoked, as in GR, although new couplings to the extra degrees of freedom could also be introduced (see \cite{Blas:2009yd,Germani:2009yt,no1} for this type of constructions). Here, we will follow previous discussions where a very general approach is taken. We will add to the gravity field equations a cosmological stress-energy tensor, such that in the low-energy limit we recover the usual GR formulation. This tensor is a hydrodynamical approximation with two quantities, $\rho,p$, corresponding to the density and pressure, and where both matter fields are related by the usual equation of state $p=w\rho$. Since one of our goals is to investigate the relation between HL and dark energy (cosmic acceleration), we will only consider here $w>0$, so that no dark energy is introduced by hand in the model.

\section{ Ho\v{r}ava-Lifshitz cosmology: detailed balance case}
\label{2}

From the discussion of the previous section, we can write the relevant field equations on a FLRW ansatz. We found, after some trivial algebra, that the system can be written as follows
\numparts
\begin{eqnarray}
\fl &&\alpha  (3 \lambda -1) \dot{H} +\alpha  (3  \lambda -1) H^2=-\frac{1}{6}
(1+3w) \rho+\mathds{A} \alpha  (1-3 \lambda ) \frac{k^2}{a^4}+\mathds{A}
\alpha  (3 \lambda -1) \Lambda ^2 \;,\label{HLComsEq1}\\
\fl&&3\alpha  (3 \lambda -1) H^2+6 \mathds{A} \alpha  (3 \lambda-1 ) \Lambda
 \frac{ k}{a^2}=\rho+3 \mathds{A} \alpha  (3 \lambda -1)\frac{ k^2}{a^4}+3
 \mathds{A} \alpha  (3 \lambda -1) \Lambda ^2 \;,\label{FriedDB}\\
\fl&&\dot{\rho}+3 (w +1) H \rho=0 \label{HLComsEq3}\;,
\end{eqnarray}
\endnumparts
where we have chosen to leave the parameters $\alpha,\lambda,\Lambda$ explicit because of their cosmological relevance, and we have defined the variable $\mathds{A}=\frac{-\zeta}{\alpha(1-3\lambda)^2}$ which is always positive, since $\zeta \leq 0$ owing to the detailed balance constraint. Notice that, already at this level, only a subset of all the initially defined couplings ($\alpha,\lambda,\Lambda,\mathds{A}$), plays a role in the cosmology.

\subsection{Dynamical analysis}
In order to analyze the phase space of  this cosmological model, let us define the variables
\begin{equation}\label{DBVar}
\Omega =\frac{\rho }{3 \alpha H^2}\ ,\qquad z=\frac{\mathds{A} \Lambda ^2}{H^2}\ ,
\qquad K=2 \Lambda\frac{ k
   \mathds{A}  }{a^2 H^2}\ , \qquad C=\frac{k^2 \mathds{A}}{a^4 H^2}\ ,
\end{equation}
with the cosmic time $\mathcal{N}=\ln[a(t)]$.

The cosmological equations \rf{HLComsEq1}-\rf{HLComsEq3} are then equivalent to the system
\numparts\label{HLDynSys}
\begin{eqnarray}
&\Omega '=\frac{1+3 w}{3 \lambda-1 }\Omega ^2+ \Omega (2 C-2 z -3 w -1)  \ ,\\
&z'=-2 z^2+z\left[2+2 C +\frac{1+3 w  }{3   \lambda-1 }\Omega\right] \, ,\\
&K'=K \left[2 C-2 z +\frac{1+ 3 w }{ 3   \lambda-1} \Omega \right]\, , \\
&C'=2 C^2-C\left[2+2 z-\frac{1+3  w }{3   \lambda-1}\Omega\right]\, ,
\end{eqnarray}
\endnumparts
with the Gauss constraint
\begin{equation}\label{GaussConstr}
1+K-z-C +\frac{\Omega }{1-3 \lambda }=0\, ,
\end{equation}
where the prime denotes the derivative with respect to $\mathcal{N}$.
The \rf{GaussConstr} allows to eliminate one of the equations. If we choose to eliminate
the one for $\Omega$  we obtain
\numparts
\begin{eqnarray}\label{HLDynSysRED}
&z'=z [3+C+K-3 z-3 w (C-K+z-1) ]\, ,\label{HLDynSysRED1}\\
&C'=C \left[C+K-3 z-3 w(C-K+z-1) -1\right]\, ,\\
 &K'=K \left[1+C+K-3 z-3 w(C-K+z-1)\right]\label{HLDynSysRED3}\, .
\end{eqnarray}
 \endnumparts
This system presents three invariant submanifolds $z=0$, $K=0$, $C=0$ which, by definition,
cannot be crossed by any orbit. This implies that no global attractor can exist in this
type of HL cosmology. Also, the structure of
\rf{GaussConstr} reveals that the system is non-compact and asymptotic analysis will be required
in order to complete the study of the phase space. Due to its dimensionality  it is not easy to draw a plot of the  phase space. Naturally, since invariant submanifolds are present, one could think to give a plot of orbits belonging to them, however the physical meaning of such manifold is not clear. For example, the invariant submanifold $C=0$ might be associated at first glance to flat cosmology, however from the definitions \rf{DBVar}, one can see that this would be the case only for $K=0$, so that the rest of  $C=0$ has no real physical meaning. Similar arguments can be given for $z=0$ and $K=0$ submanifolds. As consequence  only in orbits in the phase space bulk correspond to physical evolutions for the system and we have describe them without the aid of graphics \footnote{Of course one could consider the $\Omega=0$ invariant submanifold which could be of interest, but in our setting this would require the derivation of a new system of the type \rf{HLDynSysRED1}- \rf{HLDynSysRED3}  in which another variable is eliminated. Since  here we are mainly interested in the role of the HL corrections, drawing $\Omega=0$ will not add much to the understanding of the dynamics of the cosmology.}.
\subsection{Finite analysis.}
Let us start with the finite analysis. The finite fixed points can be found setting $z',C',K'$  in
\rf{HLDynSysRED1}-\rf{HLDynSysRED3} to be zero and solving the corresponding algebraic equations. The results are shown
in Table \ref{PF-Fin}. The solutions associated to the fixed point can be derived from the
Raychaudhuri equation
\begin{equation}
\dot{H}=\frac{1}{2} \left[3 z-C-K -3+3 w(C-K+z-1) \right] H^2\;,
\end{equation}
 and the results are shown in Tab.~\ref{PF-Fin}, too. As one can see there, we have a general
 Friedmann solution which depends on the barotropic factor $w$  of standard matter,
 a pure radiation like solution, a Milne universe, and an exponential solution.
 Note that the Friedmann solution is generated when the term related to the spatial
 curvature $k$ is dominant. Instead, when the effective Ho\v{r}ava radiation is
 dominant, the corresponding cosmology presents a ``radiation-dominated like" solution, as expected
 from the form of the corresponding terms in \rf{HLComsEq1}-\rf{HLComsEq3}. Substitution into
 Eqs.~\rf{HLComsEq1}-\rf{HLComsEq3}  reveals that this is actually a vacuum solution and that the
 exponential solution corresponds to a an oscillating evolution with period
 $T=2\pi\frac{3\lambda-1}{\Lambda} \sqrt{\frac{\alpha}{|\zeta|}}$. Such solution can
 be connected to the scenario proposed in \cite{Brandenberger:2009yt}.
 The Milne solution, instead,  is not an actual solution of the
 system\footnote{However this does not constitute a real problem because, as we
 will see, this point is always unstable and there is no orbit which can reach it.}.

The stability of the fixed points can be determined by evaluating the eigenvalues of the Jacobian
matrix  associated with the system \rf{HLDynSysRED1}-\rf{HLDynSysRED3}, as prescribed by the Hartman-Grobman Theorem
\cite{Hartman}. The results can be found in Table \ref{PF-Fin}. We can observe that the
thermodynamical properties of matter influence the stability of the Friedmann point  $\mathcal{A}$
and of the point $\mathcal{B}$. This means that, if $0<w<1/3$, the typical completely finite orbit
implies an initial radiation-like  behavior that evolves towards a Friedmann or a Milne behavior (or
both), to eventually approach an oscillating state. Instead, if $1/3<w<1$, $\mathcal{A}$  is a
source so that the typical orbit will start with a Friedmann evolution and evolve towards a
radiation-like or a Milne evolution before converging to an oscillatory behavior. In  both these
scenarios we {\it do not} find any transition to a dark energy era.

In fact, in terms of the phase space one can characterize this transition as the fact that  the
orbits will cross the plane
\begin{equation}
C+K-3 z-3 (C-K+z-1)w +1=0\;,
\end{equation}
but none of the finite orbits described above can actually achieve that. This conclusion was in some sense expected, since it was already noticed \cite{Nastase:2009nk} that in this case the cosmological term
seems to have the wrong sign when the detailed balance condition is imposed. 

\begin{table}
\caption{Finite fixed points of the system \rf{HLDynSysRED1}-\rf{HLDynSysRED3} and their associated solutions.}\label{PF-Fin}
\begin{center}
\begin{tabular}{lclll}
\hline\hline Point& Coordinates  & Solution& Energy density& Stability\\
&$[\Omega,z, C, K]$&&\\
\hline\\
$\mathcal{A}$&  $[3\lambda-1,0, 0, 0]$ & $a= a_0 (t-t_0)^{\frac{2}{3(1+w)}}$&$\rho=
\rho_0(t-t_0)^{-2}$&$
\left\{
\begin{array}{ll}
\mbox{saddle}  & 0\leq w\leq1/3  \\
 \mbox{repeller}& 1/3\leq w\leq1 \\
\end{array}
\right.
$\\
\\
$\mathcal{B}$&  $\left[0, 0,1, 0\right]$ & $a= a_0 (t-t_0)^{\frac{1}{2}}$&$\rho=0$&  $
\left\{
\begin{array}{ll}
\mbox{repeller} & 0\leq w\leq1/3  \\
\mbox{saddle} & 1/3\leq w\leq1 \\
\end{array}
\right.
$
\\
\\
$\mathcal{C}$&   $\left[0, 0, 0, -1\right]$ & $a= a_0 (t-t_0)$&$\rho=0$& saddle\\
\\
$\mathcal{D}$& $\left[0,1, 0, 0\right]$  & $a=a_0 e^ {\tau (t-t_0)}$&$\rho=0$&  attractor\\
\\
\hline
\\
\multicolumn{5}{c}{ $\tau=i\sqrt{\frac{|\zeta|}{\alpha}}\frac{\Lambda}{3\lambda-1}$}\\
\\
\hline\hline\\
\end{tabular}\end{center}
\end{table}

\subsection{Asymptotic analysis}

Owing to the fact that the dynamical system (\ref{HLDynSys}) is non-compact, there could be features
in the asymptotic regime  which are non trivial for the global dynamics. Thus, in order to complete
the analysis of the phase space we will now extend our study using the Poincar\'{e} projection method
\cite{Lefschetz}. This method is based on a phase space coordinate change
\begin{equation}\label{PoincTransf}
C=\rho\sin \theta \cos \phi, \qquad K=\rho\sin \theta \sin \phi, \qquad z=\rho \cos \theta\,,
\end{equation}
where
\begin{equation}
\rho=\frac{r}{1+r}\ ,
\end{equation}
$\theta \in [0,\pi]$, and  $\phi \in [0, 2\pi]$.

In this way, the limit $\rho\rightarrow 1$ corresponds to $r \rightarrow \infty$, and one can study
the asymptotics of the system. Performing the transformation \rf{PoincTransf}, the system
\rf{HLDynSysRED} becomes, in the limit  $\rho\rightarrow 1$,
\numparts
\begin{eqnarray}\label{HLDynSysAsy}
&\rho'\rightarrow 3
(1+w) \cos\theta-\sin\theta \left[(1-3w)  \cos\phi+(1+3 w)\sin\phi\right]\ ,\label{eqRho}\\
&\theta'\rightarrow\frac{1}{2}\sin(2\theta) [3+\cos (2 \phi )] \ ,\\
&\phi'\rightarrow\sin (2 \phi)\ .
\end{eqnarray}
\endnumparts
Notice that the radial equation does not contain the
radial coordinate, so that the fixed points can be obtained using
 the angular equations only. Setting $\phi '=0$ and $\theta '=0$, we
obtain the fixed points which are listed in Table \ref{PF-Asy}. The solutions associated with these
points can be found with the method described in \cite{Carloni:2004kp}. The corresponding results
are also given in Table \ref{PF-Asy}.

The stability of these points is studied by analyzing first the stability of the angular coordinates
and then deducing, from the sign of Eq.~\rf{eqRho}, the stability on the radial direction
\cite{Carloni:2004kp}. Ensuing results are given in Table \ref{Stab-Asy}. The presence of these
additional points greatly enriches the dynamics of the cosmology. The new points represent a Gauss
function evolution that is characterized by a very fast expansion and a similarly fast recollapse.
If unstable, such points could represent---at least in principle---an inflationary or dark energy
era. However, none of the orbits connecting these points can be considered of much cosmological
relevance, because the ones that involve the asymptotic fixed points with the right stability
behavior do not contain a transient Friedmann era.

\begin{table}
\begin{center}
\begin{tabular}{lcc}
\hline\hline Point& Coordinates  &  Solution\\
&$(\phi,\theta)$&\\
\hline\\
$\mathcal{E}_1$&  $[0, 0]$ & $a= a_0e^{\frac{3}{2} (w +1) \left(t-t_0\right){}^2} $\\
\\
$\mathcal{E}_2$&  $[\pi, \pi]$ & $a= a_0e^{\frac{3}{2} (w +1) \left(t-t_0\right){}^2} $\\
\\

$\mathcal{F}_1$&  $\left[0, \pi/2\right]$ & $a= a_0 e^{\frac{1}{4} (1-3 w )
   \left(t-t_0\right){}^2}$\\
\\
$\mathcal{F}_2$&  $\left[\pi, \pi/2\right]$ & $a= a_0 e^{\frac{1}{4} (1-3 w )
   \left(t-t_0\right){}^2}$\\
\\
$\mathcal{G}_1$&  $\left[\pi/2,\pi/2\right]$ & $a= a_0 e^{ \frac{1}{4} (1+3 w )
   \left(t-t_0\right){}^2}$\\
\\
$\mathcal{G}_2$&  $\left[3\pi/2,\pi/2\right]$ & $a= a_0 e^{\frac{1}{4} (1+3 w )
   \left(t-t_0\right){}^2}$\\ 
 \\
 \hline\hline
\end{tabular}\caption{Asymptotic fixed points of the system \rf{HLDynSysRED1}-\rf{HLDynSysRED3}, their associated solutions, and
stability.}\label{PF-Asy}
\end{center}
\end{table}

\begin{table}
\caption{Stability of the asymptotic fixed points of the system \rf{HLDynSysRED1}-\rf{HLDynSysRED3}.}\label{Stab-Asy}
\begin{center}
\begin{tabular}{llcl}
\hline\hline Point& \ \ Eigenvalues  & $\rho'$& Stability\\
\hline\\
$\mathcal{E}_1$& $\left[2i\sqrt{2},-2i\sqrt{2}\right]$ & $\rho'>0$ always& attractive center\\
\\
$\mathcal{E}_2$& $\left[2i\sqrt{2},-2i\sqrt{2}\right]$ &  $\rho'<0$ always & repulsive center\\
\\
$\mathcal{F}_1$&   $\left[2\sqrt{2},-2\sqrt{2}\right]$ &$
\left\{
\begin{array}{ll}
\rho'>0 &0\leq w <\frac{1}{3},  \\
\rho'<0 &  \frac{1}{3}<w \leq 1\\
\end{array}
\right.
$&saddle\\
\\
$\mathcal{F}_2$&   $\left[2\sqrt{2},-2\sqrt{2}\right]$ &$
\left\{
\begin{array}{ll}
\rho'>0  &\frac{1}{3}<w \leq 1,  \\
\rho'<0 & 0\leq w<\frac{1}{3},\\
\end{array}
\right.
$&saddle\\
\\
$\mathcal{G}_1$& $\left[2 i,-2i\right]$ &  $\rho'<0$ always& repulsive center\\
\\
$\mathcal{G}_2$&  $\left[2i,-2i\right]$ &  $\rho'>0$ always& attractive center \\
\\
\hline\hline
\end{tabular}
\end{center}
\end{table}
\section{ Ho\v{r}ava-Lifshitz cosmology: no detailed balance case (SVW potential)}
\label{3}
If we do not impose detailed balance to hold, the cosmological equations in presence of matter
(assumed to be a barotropic fluid) can be written as  (we closely follow the notation of \cite{Sotiriou:2009bx})
\numparts
\begin{eqnarray}
&\left(1-\frac{3 \xi }{2}\right)\left( \dot{H}+H^2\right)+\frac{1}{2}
\kappa ^2\rho(1+3 w) -\frac{\chi_1}{6}+\frac{\chi _3 k^2}{6 a^4}+
\frac{\chi _4 k}{3 a^6} =0\;,\label{HLComsEqNoDB1}\\
&\left(1-\frac{3 \xi }{2}\right)H^2-\frac{\chi _2  k}{6 a^2}-\kappa ^2
\rho-\frac{\chi _1}{6}-\frac{\chi _3 k^2}{6 a^4}-\frac{\chi_4 k}{6 a^6} =0
\;,\label{FriedNoDB}\\
&\dot{\rho}+3 (w +1) H \rho=0 \label{HLComsEqNoDB3}\;,
\end{eqnarray}
\endnumparts
where
\numparts
\begin{eqnarray}
&\kappa ^2=\frac{1}{6\alpha}\;,\\
&\chi_1=\frac{g_0 \alpha^3}{6}\;,\\
&\chi_2=- 6 g_1\alpha^2>0\;,\\
&\chi_3=12\alpha\left(3 g_2+g_3\right)\;,\\
&\chi_4=24 \left(9 g_4+3 g_5+g_6\right)\;,
\end{eqnarray}
\endnumparts
and we have defined $\xi=1-\lambda$ assuming also that $\xi\neq 2/3$. Notice that taking $\xi>2/3$ would imply that the energy density in the Friedmann equation has a negative sign. In fact, this range of the parameter corresponds to spin zero modes of the theory, which can be excluded at the cosmological level and are related to unwanted ghost modes. Also the sign of the term $\chi_1$ determines the sign of an (effective) cosmological constant in the model and $\chi_2$ can be always taken to be negative \cite{Sotiriou:2009bx}. Comparing the system above with the one in \rf{HLComsEq1}-\rf{HLComsEq3} one can note that, as pointed out in \cite{Sotiriou:2009bx}, there is not much
difference between the two cases. For example, apart from the values of the constants, \rf{FriedDB}
and \rf{FriedNoDB} differ only by the term associated with $\chi_4$. However, as we are going to show,
the associated cosmological dynamics will be non-trivially changed.

\subsection{Dynamical analysis}
Let us define the variables
\begin{equation}\label{VarDynNoDB}
\fl\Omega =\frac{\kappa ^2 \rho }{H^2},\qquad x=\frac{k^2 \chi_3}{6 a^4 H^2}, \qquad y=\frac{k
\chi_4}{6 a^6 H^2}, \qquad z=\frac{\chi_1}{6 H^2}, \qquad K=\frac{k \chi_2}{6 a^2 H^2}\ ,
\end{equation}
and the cosmic time $\mathcal{N}=\ln[a(t)]$. The cosmological equations \rf{HLComsEqNoDB1}-\rf{HLComsEqNoDB3} are then
equivalent to the system
\numparts
\begin{eqnarray}
&\Omega '=\left[\frac{4( z-x-2 y)}{2-3\xi}-1-3 w\right]\Omega-\frac{2(1+3 w)}{2-3\xi} \Omega ^2 \ ,\label{HLDynSysNoDB1}\\
& x'=\frac{2 x}{2-3\xi} [-4 y+2 z-(1+3 w) \Omega  -2]-\frac{4 x^2}{2-3\xi}  \ ,\\
&y'=\frac{2 y}{2-3\xi} [-2 x+2 z-(1+3 w) \Omega-4]-\frac{8 y^2}{2-3\xi} \ , \\
&z'=\frac{2 z}{2-3\xi}[-2 x-4 y-(1+3 w) \Omega +2] +\frac{4 z^2}{2-3\xi} \ ,\\
&K'=-\frac{2 K}{2-3\xi} [2 x+4 y-2 z+(1+3 w) \Omega ]\ ,\label{HLDynSysNoDB4}
\end{eqnarray}
\endnumparts
with the Gauss constraint
\begin{equation}\label{GaussConstr2}
K+x+y+z+\Omega +1-\frac{3}{2}\xi=0,
\end{equation}
where the prime denotes derivation with respect to $\mathcal{N}$. As before, the constraint
\rf{GaussConstr} allows to eliminate one of the equations. If we choose to eliminate the one for
$\Omega$, we obtain
\numparts
\begin{eqnarray}
\fl& x'=\frac{2(1-3 w) x^2}{2-3 \xi }\nn\\
\fl&\quad+\frac{x}{2-3 \xi } [2 K (3 w+1)-6 (w-1) y-6 (w+1) z-(1-3 w) (2-3
   \xi )] \label{HLDynSysREDNoDB1}
   \ ,\\
\fl&y'=-\frac{6 (w-1) y^2}{2-3 \xi }\nn\\
\fl&\quad+\frac{y}{2-3 \xi } [2 K (3 w+1)+2 (1-3 w) x-6 (w+1) z+3 (w-1)
   (2-3 \xi)]
   \ , \\
\fl&z'=-\frac{6 (w+1) z^2}{2-3 \xi }\nn\\
\fl&\quad+\frac{z}{2-3 \xi } [2 K (3 w+1)+2 (1-3 w) x-6 (w-1) y+3 (w+1)
   (2-3 \xi )]
\ ,\\
\fl&K'=\frac{2 K^2 (3 w+1)}{2-3 \xi }\nn\\
\fl&\quad+\frac{K}{2-3 \xi } [2 (1-3 w) x-6 (w-1) y-6 (w+1) z+(3 w+1) (2-3
   \xi )].\label{HLDynSysREDNoDB4}
\end{eqnarray}
\endnumparts
As expected, the new degrees of freedom, associated with additional terms present in this case, result
in an additional dimension for the phase space. The system above possesses four invariant
submanifolds, namely $x=0$, $y=0$, $z=0$, $K=0$. This implies that, also in this case, no global
attractor can exist. As before, the structure of  \rf{GaussConstr2} is such that the system  is
non-compact and this will require an asymptotic analysis to be performed. Also in this case, the dimensionality of the phase space and the nature of the invariant submanifolds is such that we cannot picture clearly the phase space. Thus we will not be able to use graphics to show our results.
\subsection{Finite analysis}
The finite fixed points are found by setting $x',y,z',K'=0$  in \rf{HLDynSysREDNoDB1}-\rf{HLDynSysREDNoDB4} and solving the
resulting  system of algebraic equations. The results are shown in Table \ref{PF-Fin-NoDB}. Note
that the presence of a fixed point depends on the exact sign of the constant $\chi_i$, as well as
on the value of $\xi$. For example, given the fact that the variable $\Omega$ is defined positive, the fixed point $\mathcal{A}$, can exist only if its coordinates are non negative, and this happens
 for $\xi<\frac{2}{3}$ only.

The solutions associated to the fixed point can be derived from the Raychaudhuri equation
\begin{equation}
\fl\dot{H}=-\frac{H^2}{4-6 \xi } \left[2 K+2 x+6 y-6 z-6 w (x+y+z-1-K)-9 (w+1) \xi +6\right].
\end{equation}
The corresponding results are shown in Table \rf{PF-Fin-NoDB}. The new terms in  the system
\rf{HLComsEqNoDB1}-\rf{HLComsEqNoDB3} induce a new fixed point characterized by a behavior $t^{1/3}$ which corresponds to the domination of a new cosmic component which goes like $a^{-6}$.  Substitution into
Eqs.~\rf{HLComsEq1}-\rf{HLComsEq3}  reveals that, in this case, only the  Friedmann solution and  the exponential
solution yield identities. Specifically, the exponential solution represents a de Sitter solution if
$\frac{\chi _1}{2-3 \xi }>0$, otherwise it is associated with oscillations. In other words, if one
wants standard matter to interact with HL gravity in the standard way (gravity makes matter to
attract itself), then the de Sitter solution is only present if $\chi_1>0$, i.e., if the
cosmological constant has the right sign, as expected.

In the same way as in the previous case, the stability of the fixed points can be determined by
evaluating the eigenvalues of the Jacobian matrix  associated with the system \rf{HLDynSysRED}, as
prescribed by the Hartman-Grobman theorem \cite{Hartman}. The results can be also found in Table
\ref{PF-Fin-NoDB}. The stabilities of the fixed points in this model are different from the
corresponding ones in the previous case although, again, the only stable finite fixed point
continues to be the de Sitter one.

As before we can characterize the transition form decelerated and accelerate expansion as the crossing of the hyperplane
\begin{equation}
 0=2 K+2 x+6 y-6 z-6 w (x+y+z-1-K)-3 (1+3w) \xi +2\,.
\end{equation}

Unfortunately, due to the higher dimensionality of the phase space, the dynamics of this model are
not as easy to extract as the ones in the previous paragraph. However in the general structure of the
equations there is no feature which prevents the existence of an orbit connecting the unstable
Friedmann phase with the de Sitter attractor. This means that HL cosmologies without detailed
balance can actually admit a transition between Friedmann evolution and a dark energy era.

\begin{table}
\begin{center}
\begin{tabular}{lclll}
\hline\hline Point& Coordinates  & Solution& Energy density& Stability\\
&$[\Omega,x,y,z, K]$&&\\
\hline\\
$\mathcal{A}$&  $\left[\frac{1}{2}(2-3\xi),0, 0, 0,0\right]$ & $a= a_0
(t-t_0)^{\frac{2}{3(1+w)}}$&$\rho=\rho_0(t-t_0)^{-2}$&  saddle\\
\\
$\mathcal{B}$&  $\left[0, \frac{1}{2}(2-3\xi), 0, 0, 0\right]$ & $a= a_0
(t-t_0)^{\frac{1}{2}}$&$\rho=0$&  saddle\\
\\
$\mathcal{C}$&  $\left[0, 0, 0, 0, \frac{1}{2}(3\xi-2)\right]$ & $a= a_0
 (t-t_0)$&$\rho=0$&  saddle\\
\\
$\mathcal{D}$&  $\left[0, 0, 0, \frac{1}{2}(2-3\xi), 0\right]$ & $a=a_0
e^ {\tau (t-t_0)}$&$\rho=0$&  attractor\\
\\
$\mathcal{E}$&  $\left[0, 0, \frac{1}{2}(2-3\xi), 0, 0\right]$ & $a= a_0
(t-t_0)^{\frac{1}{3}}$&$\rho=0$&  repeller\\
\\
\hline
\\
\multicolumn{5}{c}{ $\tau=\sqrt{\frac{\alpha _1}{3(2-3 \xi) }}$}\\
\\
\hline\hline\\
\end{tabular}\caption{Finite fixed points of the system \rf{HLDynSysREDNoDB1}-\rf{HLDynSysREDNoDB4}, their associated solutions and stability.}\label{PF-Fin-NoDB}
\end{center}
\end{table}


\subsection{Asymptotic analysis}
Let us now look at the asymptotic analysis of the dynamical system \rf{HLDynSysNoDB1}-\rf{HLDynSysNoDB4} which, like
\rf{HLDynSys}, is non-compact. We will use again the Poincar\'{e} projection method \cite{Lefschetz}.
Let us perform the phase space coordinate change
\begin{equation}\label{PoincTransf2}
\fl x= \rho \cos\phi  \sin\theta \sin\psi,\quad y= \rho \sin\theta \sin\phi  \sin\psi, \quad z=
\cos\theta \sin\psi, \quad K= \rho\cos\psi\,,
\end{equation}
where again
\begin{equation}
\rho=\frac{r}{1+r}\ ,
\end{equation}
$\theta \in [0,\pi]$, and  $\phi \in [0, 2\pi]$. Performing the transformation \rf{PoincTransf2}, the
system \rf{HLDynSysREDNoDB1}-\rf{HLDynSysREDNoDB4} becomes now, in the limit  $\rho\rightarrow 1$,
\begin{eqnarray}\label{HLDynSysAsy2}
\numparts
&\nn\rho'\rightarrow \frac{2}{2-3\xi}\{\sin \theta [(1-3 w) \cos \phi+3 (1-w) \sin \phi]-3 (w+1)
 \cos \theta\} \sin\psi
\\&-\frac{2(3 w+1)}{2-3\xi} \cos \psi\ ,\label{eqRho2}\\
&\theta'\rightarrow\frac{1}{2}\sin(2\theta) [\cos (2 \phi )-5] \ ,\\
&\phi'\rightarrow\sin (2 \phi)\ , \\
& \psi '\rightarrow-\frac{1}{2}\sin(2\psi) \left[2 \cos ^2\theta+(\cos (2 \phi )-3) \sin
^2\theta\right] .
\endnumparts
\end{eqnarray}
Also here the radial equation does not contain the radial coordinate, so that the fixed points can
be obtained using just the angular equations. Setting $\phi '=0$, $\theta '=0$ and $ \psi '=0$, we
obtain the fixed points listed, together with their solutions, in Table \ref{PF-AsyNoDB}. Their
stability behavior is given in Table \ref{PF-Stab-AsyNoDB}.

\begin{table}
\begin{center}
\begin{tabular}{lcl}
\hline\hline Point& Coordinates  & \ \ \ Solution\\
&$[\phi,\theta.\psi]$&\\
\hline\\
$\mathcal{F}_1$&  $[0, 0,0]$ & $a= a_0e^{\frac{1}{2} (1+3w ) \left(t-t_0\right){}^2} $\\
\\
$\mathcal{F}_2$&  $[\pi, \pi,0]$ & $a= a_0e^{\frac{1}{2} (w +1) \left(t-t_0\right){}^2} $\\
\\

$\mathcal{G}_1$&  $\left[0,0, \pi/2\right]$ & $a= a_0 e^{\frac{3}{2} (w +1)
   \left(t-t_0\right){}^2}$\\
\\
$\mathcal{G}_2$&  $\left[0,\pi, \pi/2\right]$ & $a= a_0 e^{\frac{3}{2} (w +1)
   \left(t-t_0\right){}^2}$\\
\\
$\mathcal{H}_1$&  $\left[\pi/2,\pi/2,\pi/2\right]$ & $a= a_0 e^{ \frac{3}{2} (w +1)
   \left(t-t_0\right){}^2}$\\
\\
$\mathcal{H}_2$&  $\left[\pi/2,\pi/2,3\pi/2\right]$ & $a= a_0 e^{\frac{3}{2} (w +1)
   \left(t-t_0\right){}^2}$\\
\\
$\mathcal{I}_1$&  $\left[0,\pi/2,\pi/2\right]$ & $a= a_0 e^{ \frac{1}{2} (1-3 w )
   \left(t-t_0\right){}^2}$\\
\\
$\mathcal{I}_2$&  $\left[\pi, \pi/2,\pi/2\right]$ & $a= a_0 e^{\frac{1}{2} (1-3 w )
   \left(t-t_0\right){}^2}$\\
  \\
\hline\hline\\
\end{tabular}\caption{Asymptotic fixed points of the system \rf{HLDynSysREDNoDB1}-\rf{HLDynSysREDNoDB4} and their associated solutions.\label{PF-AsyNoDB}}
\end{center}
\end{table}

\begin{table}
\begin{center}
\begin{tabular}{llcc}
\hline\hline Point& \ \ Eigenvalues  & $\rho'$& Stability\\
\hline\\
$\mathcal{F}_1$& $\left[2\sqrt{2},-2\sqrt{2},-2\right]$ & $\rho'<0$ &saddle
\\
\\
$\mathcal{F}_2$& $\left[2\sqrt{2},-2\sqrt{2},-2\right]$ &  $\rho'>0$  & saddle
\\
\\
$\mathcal{G}_1$&   $\left[2\sqrt{2},-2\sqrt{2},2\right]$ & $\rho'<0$&saddle
\\
\\
$\mathcal{G}_2$&    $\left[2\sqrt{2},-2\sqrt{2},2\right]$ & $\rho'>0$&saddle\\
\\
$\mathcal{H}_1$& $\left[-4, -2 \sqrt{3}, 2 \sqrt{3}\right]$ &  $\rho'<0$ & saddle\\
\\
$\mathcal{H}_2$& $\left[-4, -2 \sqrt{3}, 2 \sqrt{3}\right]$ &  $\rho'>0$ &saddle \\
\\
$\mathcal{I}_1$& $\left[2i\sqrt{2},-2i\sqrt{2},-2\right]$ &  $\rho'<0$ & $
\left\{
\begin{array}{cc}
 \xi<\frac{2}{3} \quad 0<w<\frac{1}{3}&  \mbox{attractive center}      \\
 \xi<\frac{2}{3} \quad \frac{1}{3}<w<1 &  \mbox{repulsive center}       \\
  \xi>\frac{2}{3} \quad 0<w<\frac{1}{3}&   \mbox{repulsive center}    \\
 \xi>\frac{2}{3} \quad \frac{1}{3}<w<1 & \mbox{attractive center}      \\
 \end{array}
\right.
$
\\
\\
$\mathcal{I}_2$&  $\left[2i\sqrt{2},-2i\sqrt{2},-2\right]$ &  $\rho'>0$& $
\left\{
\begin{array}{cc}
 \xi<\frac{2}{3} \quad 0<w<\frac{1}{3}&  \mbox{repulsive center}      \\
 \xi<\frac{2}{3} \quad \frac{1}{3}<w<1 &  \mbox{attractive center}       \\
  \xi>\frac{2}{3} \quad 0<w<\frac{1}{3}&   \mbox{attractive center}    \\
 \xi>\frac{2}{3} \quad \frac{1}{3}<w<1 &     \mbox{repulsive center}   \\
\end{array}
\right.
$
 \\
\hline\hline\\
\end{tabular}\caption{Stability of the asymptotic fixed points of the system \rf{HLDynSysREDNoDB1}-\rf{HLDynSysREDNoDB4}.\label{PF-Stab-AsyNoDB}}
\end{center}
\end{table}

\section{Discussion and conclusions}
\label{4}
In this paper we have used dynamical system techniques to analyze the non vacuum cosmology of
Hor\^{a}va-Lifshitz gravity both in the presence and in the absence of detailed balance. Our analysis has allowed to both gain an understanding of the qualitative behavior of the cosmology and to obtain interesting new constraints on the parameters of the model.

In the first
case the phase space exhibits four finite fixed points, three of which represent physical solutions
of the system. Although one of these points is associated to an unstable classical Friedmann
solution that could be certainly useful to model the nucleosynthesis and the structure formation
periods, our analysis does not reveal any useful fixed point which could model an inflationary or
dark energy phase. It has been proposed that, because of the changes in the value of the speed of
light contained in the theory, the absence of an explicit inflationary phase might not be such a serious
shortcoming of the theory \cite{Kiritsis:2009sh}, although at first glance it is difficult in this setting to produce a dark energy era. A more conclusive analysis of this issue, however, will require a complete
numerical study, which is left to future work.

An interesting result of our investigation is that one of the attractors in the theory corresponds
to an oscillatory behavior. Such oscillations can be associated with a bouncing universe, which can be
connected to the analysis in \cite{Brandenberger:2009yt}.
Other attractors are associated with a recollapsing solution in the form of a Gauss function.  These
solutions, which also appear in the context of GR plus a scalar field,  higher-order and
scalar-tensor gravities, could---if proven to be unstable---model some kind of superinflationary
phase whose properties ought to be carefully investigated. However, most of the orbits that contain
these points will either evolve towards recollapse or to the oscillatory attractor. In other words, when detailed balance is imposed, although one is able to prove that the corresponding cosmology could, given suitable initial conditions, behave like a Friedmann one, it turns out that it will eventually evolve towards an oscillatory behavior or either recollapse, leaving no space for a dark-energy era.

Maybe the most important result in this paper is related to the HL cosmology  without detailed balance. In this case, in fact, the additional freedom in the values of the parameters allows the existence of cosmic histories which contain a Friedmann era and evolve towards a dark energy one. This follows because the phase space contains  a fixed point associated to the standard Friedmann solution which is unstable and another one which can be associated to a de Sitter solution which is an attractor. The last point can then model a dark energy era. However, the existence of these fixed points is only a necessary condition: because of the presence of invariant manifolds and the constraints on the parameters only a subset of the phase space and the parameter space will realize this scenario.

On the other hand the fact that the fixed points all lie on invariant submanifolds guarantee that such orbits can exist. Unfortunately the high dimension of the phase space makes it quite hard to perform any qualitative analysis. Therefore only numerical methods will allow the investigation of the details of these orbits. Notwithstanding these problems, we feel that it is safe to conclude that a cosmology compatible with the present observations can be obtained, in the HL framework, {\it only} if the detailed balance is broken. Such result makes this type of HL gravity a very promising phenomenological model for both the study of dark energy and quantum gravity.

\ack
This work was partially funded by Ministerio de Educaci\'on y Ciencia, Spain, projects
CICYT-FEDER-FPA2005-02211 and FIS2006-02842, CSIC under the I3P program, and AGAUR, contract
2005SGR-00790 and grant 2009BE-100132. SC was funded by Generalitat de Catalunya through the Beatriu
de Pin\'{o}s contract 2007BP-B1 00136. EE's research was performed while on leave at Department of
Physics and Astronomy, Dartmouth College, 6127 Wilder Laboratory, Hanover, NH 03755, USA.

\end{document}